\documentclass[a4paper]{article}
\usepackage{ISCSLP2026}
\usepackage{ifthen}
\usepackage{hyperref}
\usepackage{multirow}
\usepackage{makecell}
\usepackage{graphicx}

\newboolean{blind}
\setboolean{blind}{false}

\title{DAVE: A Decoupled Audio-Visual Enhancement Framework for Real-World Speech Separation}

\name{
	\ifthenelse{\boolean{blind}}{Anonymous to ISCSLP}
	{Wei Zhou$^{1}$, Wanyi Ning$^{1,2,*}$\thanks{$^{*}$Corresponding author.}, Yinshang Guo$^{3}$, Qianxiao Fang$^{1}$, Haitao Qian$^{1}$, Yingpeng Li$^{1}$}
}

\address{
  \ifthenelse{\boolean{blind}}{Anonymous to ISCSLP}
  {
  	$^1$Yijiahe AI \quad
  	$^2$Tianjin University \quad
  	$^3$Nanjing University
  }
}

\email{
	\ifthenelse{\boolean{blind}}{Anonymous to ISCSLP}
	{zhouwei@yijiahe.com, ningwanyi@126.com}
}

\begin{document}

\maketitle

\begin{abstract}
Audio-visual speech enhancement under real-world conditions remains challenging due to unreliable visual inputs and the lack of large-scale training data with realistic acoustic conditions. Existing approaches usually fuse visual features directly into the separation network, making them vulnerable to degraded visual signals. 
In this paper, we present DAVE, a decoupled audio-visual enhancement framework for real-world speech separation. Firstly, to address the data scarcity issue, we construct DAVE-Corpus, a large-scale training corpus with 219,411 mixtures generated from public meeting corpora through combinatorial acoustic augmentation. Then, we introduce a progressive multi-objective optimization strategy to jointly improve speech separation, intelligibility, speaker identity preservation, and perceptual quality. We further develop a certified selective enhancement chain that applies scene routing, GAN-based denoising, and loudness normalization only within the no-reference partition, guaranteeing non-degradation of reference-based metrics. Experimental results on the Real-World Audio-Visual Speech Enhancement Challenge\footnote{\url{https://real-world-avse.github.io/}} demonstrate the robustness of DAVE under both real-world mixed scenarios and visual degradation conditions.
\end{abstract}


\section{Introduction}

Audio-visual speech enhancement (AVSE) aims to recover the speech signal of a target speaker from noisy and interfering speech by exploiting complementary acoustic and visual information. Compared with audio-only speech separation approaches~\cite{luo2019convtasnet,yu2017pit,hershey2016deepclustering,ning2026ps4proxysupervisedjointtraining}, AVSE benefits from visual cues such as facial movements and speaker appearance, which provide valuable information for identifying and extracting the target speaker~\cite{ephrat2018looking,afouras2018conversation}. Therefore, AVSE has been widely studied for applications including assistive listening, human-computer interaction, and multi-speaker communication systems. However, deploying AVSE systems in real-world environments remains challenging. Unlike conventional benchmarks based on clean videos and artificially mixed speech, real-world scenarios often involve naturally occurring speech overlap, reverberation, environmental noise, and degraded visual observations. Real-world challenges such as the CHiME-5 task~\cite{barker2018chime5} highlight the difficulty of multi-party meeting recognition under noisy, far-field conditions. These factors create significant challenges for building robust AVSE models.

Existing AVSE approaches mainly focus on designing effective audio-visual fusion mechanisms. Typically, audio and visual representations are extracted separately and then fused within the separation network, allowing visual information to directly guide speech reconstruction~\cite{wu2019avconvtasnet}. Representative separation backbones include Conv-TasNet~\cite{luo2019convtasnet}, DPRNN~\cite{luo2020dprnn}, and SepFormer~\cite{subakan2021sepformer}, which have achieved strong performance on synthetic benchmarks. Although such approaches achieve promising performance under controlled conditions, they heavily rely on the quality and reliability of visual inputs. In real-world scenarios, visual signals may be affected by occlusion, low resolution, missing frames, blur, or far-field recording conditions. When degraded visual features are directly incorporated into the separation process, inaccurate visual information may interfere with speech reconstruction and reduce system robustness. Moreover, the lack of large-scale training data with realistic acoustic conditions further limits the generalization ability of existing AVSE systems, as models trained on simulated mixtures often suffer from a considerable domain gap when applied to real-world recordings.

In this paper, we propose DAVE, a decoupled audio-visual enhancement framework for real-world speech separation. Unlike conventional fusion-based approaches, DAVE separates audio reconstruction from visual speaker attribution, where the audio branch focuses on robust speech separation and the visual branch provides speaker-related information for target identification. Furthermore, we construct DAVE-Corpus, containing 219,411 realistic mixtures generated from public meeting corpora~\cite{du2022alimeeting,wang2021misp,fu2021aishell4} through combinatorial acoustic augmentation with room impulse response simulation~\cite{scheibler2018pyroomacoustics}. Based on DAVE-Corpus, we develop a progressive multi-objective optimization strategy and a metric-preserving selective enhancement chain to improve separation quality, intelligibility, speaker identity preservation, and perceptual quality.

The main contributions of this paper are summarized as follows: (1) we construct DAVE-Corpus, a large-scale AVSE training corpus with 219,411 realistic mixtures, providing diverse acoustic conditions for robust model training; (2) we propose DAVE, a decoupled audio-visual enhancement framework with progressive multi-objective optimization and metric-preserving selective enhancement, improving robustness against unreliable visual inputs; and (3) we conduct extensive experiments on the Real-World Audio-Visual Speech Enhancement Challenge under realistic mixed scenarios and degraded visual conditions, demonstrating the effectiveness of DAVE in challenging real-world environments.

\section{Data Construction}

To train the audio separation backbone of DAVE under realistic acoustic conditions, we construct a large-scale audio training corpus, termed DAVE-Corpus, in three stages: synthetic mixture generation, official protocol remix, and EchoSet replay.

\textbf{Synthetic Mixture Generation.}
We collect speech segments from three public meeting corpora: AliMeeting~\cite{du2022alimeeting}, MISP~\cite{wang2021misp}, and AISHELL-4~\cite{fu2021aishell4}. Since the original recordings contain different levels of noise and interference, we first perform quality-based screening using frame-level energy statistics~\cite{rabiner1989theory}. Segments are retained if their estimated SNR is above 25~dB, duration is between 2.0 and 15.0~seconds, and silence ratio is below 35\%. After filtering, 4,431 speech segments are selected, including 1,469 from AliMeeting, 2,926 from MISP, and 36 from AISHELL-4.

A key challenge in real-world AVSE is the acoustic mismatch between conventional additive mixtures and real recordings. To motivate our augmentation strategy, we first analyze this mismatch using a pre-trained TIGER~\cite{xu2024tiger} separation model. As shown in Table~\ref{tab:domain_gap}, models trained with additive-only mixtures exhibit a substantial discrepancy between synthetic and real-domain performance. Further fine-tuning on additive mixtures improves synthetic performance but leads to a larger real-domain gap, suggesting that models may overfit to simplified acoustic conditions.
Based on this observation, we design a combinatorial augmentation pipeline to generate diverse meeting-like mixtures. For each mixture, two speech segments are randomly sampled while avoiding highly similar speakers. Specifically, speaker embeddings are extracted using a pretrained speaker encoder, and a pair is rejected if the cosine similarity exceeds 0.5. The mixture duration is randomly sampled between 4.0 and 8.0~seconds, and each source is truncated or repeated accordingly. The second source is randomly scaled with a relative gain sampled from $[-3,3]$~dB to simulate different speaker distances. In addition, we use a frozen Fun-ASR-Nano-2512 model~\cite{gao2023funasr} to generate reference transcriptions for all training mixtures, providing the text labels required by the CER loss during training.

\begin{table}[t]
\centering
\caption{Synthetic-real performance discrepancy analysis using the TIGER probe model. The gap is measured as the absolute difference between synthetic and real-domain performance.}
\label{tab:domain_gap}
\begin{tabular}{lccc}
\toprule
\textbf{Training Data} & 
\textbf{Synthetic} & 
\textbf{Real} &
\textbf{Absolute Gap} \\
\midrule
Additive-only 
& 9.39 
& 10.70
& 1.31 \\
Additive fine-tuned
& 14.05
& 10.49
& 3.56 \\
DAVE-Corpus
& 11.15
& 10.70
& 0.45 \\
\bottomrule
\end{tabular}
\end{table}

To model realistic acoustic environments, 50\% of the mixtures are augmented with simulated reverberation using pyroomacoustics~\cite{habets2006rir,scheibler2018pyroomacoustics}. Room configurations are randomly sampled with dimensions of $3$--$8$~m $\times$ $3$--$7$~m $\times$ $2.5$--$3.5$~m and RT60 values between 0.2 and 0.4~seconds. Each source is convolved with its corresponding room impulse response before mixing. The remaining mixtures use additive mixing only. Environmental noise from MUSAN~\cite{snyder2015musan} is further added with randomly selected SNR values between 12 and 30~dB. The final synthetic subset contains 199,998 mixtures.

\textbf{Official Protocol Remix.}
The challenge development set provides a remix partition where mixtures are created by directly adding two single-speaker waveforms following the official protocol. Using this protocol, we re-synthesize 8,000 training mixtures by randomly pairing de-duplicated single-speaker segments from the development set, while excluding 150 held-out sessions reserved for validation. This anchors the training distribution to the evaluation protocol.

\textbf{EchoSet Replay.}
The TIGER~\cite{xu2024tiger} separation model was pre-trained on the EchoSet corpus, which contains synthetic mixtures with realistic reverberation. To prevent catastrophic forgetting~\cite{kirkpatrick2017overcoming} of the pre-training domain during fine-tuning on the new DAVE-Corpus, we include 11,413 replay pairs from the EchoSet corpus as an auxiliary subset. These replay pairs are mixed into the training data alongside the synthetic and official protocol mixtures, preserving the model's ability to generalize across acoustic domains.
The final DAVE-Corpus totals 219,411 training mixtures across three subsets: 199,998 synthetic pairs bridging the synthetic-real domain gap, 8,000 official protocol remixes anchoring the training distribution to the evaluation protocol, and 11,413 EchoSet replay pairs preventing catastrophic forgetting of the pre-training domain.

\section{Methodology}
\begin{figure}[t]
\centering
\includegraphics[width=1.0\linewidth]{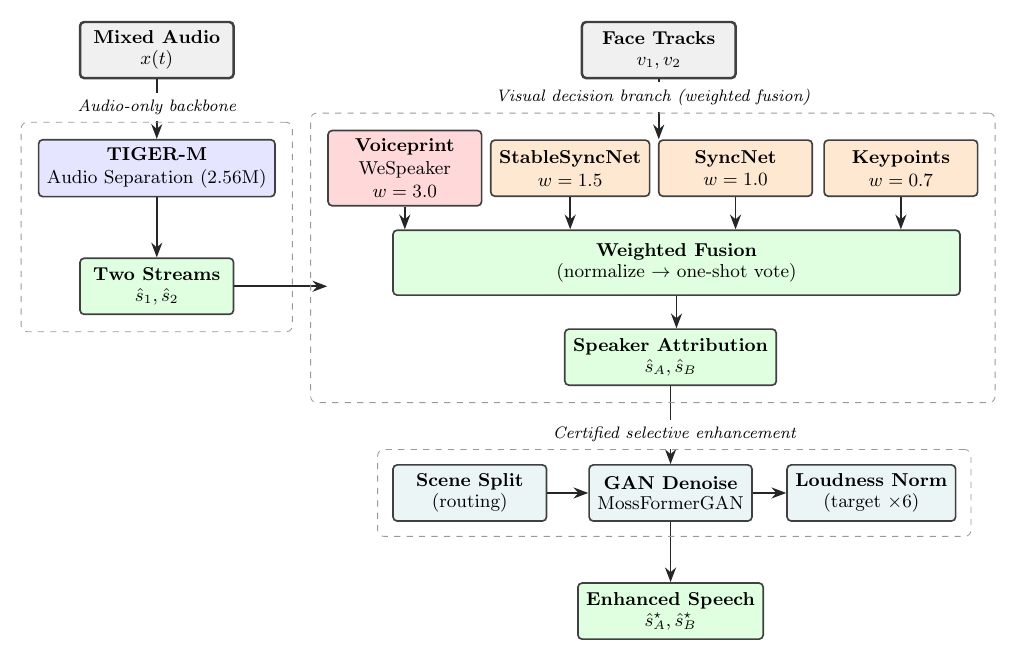}
\caption{Overview of the DAVE framework. The audio-only separation backbone reconstructs two anonymous speech streams, while a visual decision branch assigns speaker identities by weighted fusion of four evidence votes: voiceprint as the primary anchor, StableSyncNet, SyncNet, and lip-keypoint motion. The certified selective enhancement chain applies scene routing, GAN-based denoising, and loudness normalization before output.}
\label{fig:dave_framework}
\end{figure}

DAVE decouples audio reconstruction from visual speaker attribution, reserving visual information for low-bandwidth decision-making tasks where it is most reliable. As illustrated in Figure~\ref{fig:dave_framework}, the framework comprises three components. First, an audio-only separation module is trained on the DAVE-Corpus with a progressive multi-objective strategy that jointly optimizes separation quality, speech recognition accuracy, speaker identity preservation, and perceptual quality. Second, a visual speaker attribution module assigns speaker identities to the output streams by weighted fusion of audio-visual evidence, without influencing the separation process itself. Third, a certified selective enhancement module applies scene routing, GAN-based denoising, and loudness normalization only within the no-reference partition, guaranteeing non-degradation of reference-based metrics.

\subsection{Audio Separation}
\label{sec:audio_sep}

We adopt TIGER~\cite{xu2024tiger}, a lightweight time-frequency domain separation network, as the audio backbone. The original architecture with 0.82M parameters exhibits capacity saturation, where adding more training data no longer improves separation performance. We therefore scale the model to 2.56M parameters by increasing the encoder channels from 256 to 512, the feature channels from 128 to 256, and the number of repeating blocks from 8 to 12, while keeping the analysis window of 640 samples and hop size of 160 samples at 16~kHz unchanged. The resulting model, termed TIGER-M, is trained from scratch on the DAVE-Corpus.
We train TIGER-M with a multi-objective optimization strategy that jointly optimizes separation quality, speech recognition accuracy, speaker identity preservation, and perceptual quality.

\textbf{Permutation-invariant SI-SDR loss.} The primary separation loss is the negative scale-invariant signal-to-distortion ratio~\cite{leroux2019sisdr} under permutation-invariant training~\cite{yu2017pit}:

\begin{equation}
\mathcal{L}_{\text{PIT}} = \min_{\pi \in \Pi} \sum_{i} -\text{SI-SDR}(\hat{s}_{\pi(i)}, s_i),
\end{equation}
where $\Pi$ denotes the set of permutations over the two output streams and $s_i$ are the reference signals.

\textbf{CER loss.} To improve speech recognition accuracy, we minimize the teacher-forcing cross-entropy of a frozen Fun-ASR-Nano-2512 ASR model~\cite{gao2023funasr} on the separated outputs:

\begin{equation}
\mathcal{L}_{\text{cer}} = \frac{1}{2} \sum_{i} \text{CE}\big(\text{ASR}(\hat{s}_i), \mathbf{y}_i\big),
\end{equation}
where $\mathbf{y}_i$ is the reference transcription. The gradient from text-level loss is back-propagated to the waveform through a differentiable log-mel filterbank front-end. The same ASR model is used for both self-labeling the training data and official CER evaluation, ensuring that the loss only penalizes degradations that are measurable by the evaluation metric.

\textbf{Speaker fidelity loss.} To preserve speaker identity, we minimize the cosine distance between the speaker embeddings of the estimated and reference signals:

\begin{equation}
\mathcal{L}_{\text{spk}} = 1 - \frac{1}{2} \sum_{i} \cos\big(\mathbf{e}(\hat{s}_i), \mathbf{e}(s_i)\big),
\end{equation}
where $\mathbf{e}(\cdot)$ is the speaker embedding extracted by a frozen WeSpeaker ResNet34~\cite{wang2023wespeaker} model, which is the official speaker similarity evaluation model. The embedding function is made differentiable by bypassing its default inference-only wrapper.

\textbf{Perceptual losses.} We further incorporate differentiable implementations of STOI~\cite{taal2011stoi} and PESQ~\cite{rix2001pesq}, each validated against the official metric with Pearson correlations of $1.0000$ and $0.9249$, respectively. A differentiable UTMOS~\cite{saeki2022utmos} critic is implemented by reusing the official UTMOSv2 weights in a trainable-compatible forward pass, with gradient clipping and hybrid-precision safeguards. These losses are computed only on samples with clean reference signals.

\subsection{Visual Speaker Attribution}
\label{sec:visual_attr}

The audio separation module outputs two anonymous streams without speaker identities. A visual speaker attribution module assigns each stream to its corresponding speaker by weighted fusion of multiple audio-visual evidence votes, without modifying the separation outputs.

Instead of relying on a single detector, we combine four complementary evidence sources into a single weighted vote. The first is the voiceprint similarity between each separated stream and the surrounding speaker context, extracted with WeSpeaker ResNet34~\cite{wang2023wespeaker} trained on CN-Celeb~\cite{fan2020cnceleb}, the same model family used for official speaker similarity evaluation; this evidence carries the highest weight and serves as the primary anchor of the attribution. The second is the audio-visual matching score from StableSyncNet~\cite{li2024latentsync}, a stable lip-sync discriminator that provides complementary audio-visual correspondence cues. The third is the audio-visual synchrony score from a pre-trained SyncNet~\cite{chung2016syncnet}, fine-tuned following the contrastive cross-modal paradigm of Perfect Match~\cite{chung2019perfectmatch}. The fourth is a synchrony metric constructed from lip-keypoint motion energy extracted by FAN~\cite{bulat2017howfar}, which is robust to degraded pixel content. Each vote is normalized and combined with its corresponding weight, namely $3.0$ for voiceprint, $1.5$ for audio-visual matching, $1.0$ for SyncNet, and $0.7$ for keypoints, to produce a single one-shot attribution decision in a single pass.
\subsection{Certified Selective Enhancement}
\label{sec:certified}

The separated outputs after speaker attribution still retain residual noise from the original recording. As shown in Figure~\ref{fig:dave_framework}, we apply a final enhancement chain in the certified selective enhancement module, which is deliberately restricted to the no-reference partition so that reference-based metrics are never modified by construction.

\textbf{Scene Routing.} The official evaluation only measures the reference-based metrics SI-SDR, PESQ, and STOI on the synthetic remix subset, whereas the perception-oriented metrics are evaluated on all partitions. This motivates restricting any enhancement to the no-reference partition. We train a lightweight acoustic scene classifier that distinguishes remix from real-recording samples using spectral statistics, achieving $98.13\%$ accuracy on the development set. Only samples routed to the no-reference class enter the subsequent enhancement steps, while samples with ground-truth references are passed through unchanged. This yields a structural guarantee that the reference-based metrics cannot degrade.

\textbf{Layered GAN denoising.} The residual environmental noise in the no-reference outputs degrades the perception-oriented metrics. However, applying denoising indiscriminately risks distorting speech content and lowering other measured metrics. We adopt a layered GAN-based denoiser~\cite{pascual2017segan} built on MossFormerGAN~\cite{zhao2024mossformer2}, which is applied only to the routed no-reference samples and denoises the residual interferer without altering the target speech content, improving perceived quality while keeping the other scored metrics stable.

\textbf{Loudness Normalization.} Finally, we apply a loudness normalization stage that scales each output by a factor of 6$\times$ relative to the original amplitude, following the ITU-R BS.1770 loudness measurement~\cite{itu2015loudness}. The gain is taken as the smaller of the target multiple and the peak-protection bound $0.99$/peak to prevent clipping, and it is verifiably lossless with respect to the speaker-similarity metric. This final stage improves the DNSMOS-OVRL~\cite{reddy2021dnsmos} score and stabilizes loudness across samples without affecting the reference-based metrics, which are confined to the untouched remix partition.

\begin{table*}[t]
\centering
\small
\caption{Stepwise ablation study on the official development set.}
\label{tab:ablation}
\begin{tabular}{lccccccc}
\toprule
\textbf{Configuration} &
\textbf{SI-SDR}$\uparrow$ &
\textbf{PESQ}$\uparrow$ &
\textbf{STOI}$\uparrow$ &
\textbf{UTMOS}$\uparrow$ &
\textbf{DNS-OVRL}$\uparrow$ &
\textbf{CER}$\downarrow$ &
\textbf{SPK}$\uparrow$ \\
\midrule
Baseline (PIT SI-SDR) & 9.84 & 2.50 & 0.799 & 1.999 & 1.624 & 21.9\% & 0.663 \\
~~~+ CER loss & 10.31 & 2.63 & 0.810 & 1.999 & 1.631 & 19.3\% & 0.679 \\
~~~+ Speaker fidelity & 10.31 & 2.63 & 0.810 & 1.999 & 1.631 & 19.3\% & 0.697 \\
~~~+ Perceptual losses & 10.23 & 2.72 & 0.817 & 2.77 & 2.02 & 17.1\% & 0.726 \\
\midrule
~~~+ Certified enhancement & 10.23 & 2.72 & 0.817 & 2.81 & 2.12 & 17.1\% & 0.726 \\
\bottomrule
\end{tabular}
\end{table*}

\begin{table*}[t]
\centering
\small
\caption{Official challenge leaderboard results on the test set for the top 5 teams per track. Track 1: Real-World Mixed; Track 2: Visual Degradation. Rank is the mean rank across the seven metrics; lower is better. CER is reported as a fraction; lower is better.}
\label{tab:results}
\begin{tabular}{llcccccccc}
\toprule
\textbf{Track} &
\textbf{System} &
\textbf{SI-SDR}$\uparrow$ &
\textbf{PESQ}$\uparrow$ &
\textbf{STOI}$\uparrow$ &
\textbf{UTMOS}$\uparrow$ &
\textbf{DNS-OVRL}$\uparrow$ &
\textbf{CER}$\downarrow$ &
\textbf{SPK}$\uparrow$ &
\textbf{Rank}$\downarrow$ \\
\midrule
\multirow{6}{*}{Track 1}
 & Official baseline & $-5.93$ & 1.137 & 0.304 & 0.812 & 1.447 & 1.025 & 0.328 & --- \\
\cmidrule{2-10}
 & audioman & 10.70 & 2.933 & 0.841 & 2.074 & 1.832 & 0.123 & 0.749 & 2.86 \\
 & AITD & $\mathbf{12.72}$ & $\mathbf{3.000}$ & $\mathbf{0.852}$ & 2.100 & 1.697 & 0.145 & $\mathbf{0.769}$ & 2.86 \\
 & twilight & 9.16 & 2.775 & 0.825 & 2.145 & 1.944 & $\mathbf{0.135}$ & 0.742 & 3.57 \\
 & \textbf{DAVE (ours)} & 10.23 & 2.717 & 0.817 & $\mathbf{2.765}$ & $\mathbf{2.022}$ & 0.171 & 0.726 & 4.00 \\
 & SUSTechAILab & 10.42 & 2.651 & 0.823 & 1.998 & 1.770 & 0.164 & 0.737 & 5.43 \\
\midrule
\multirow{6}{*}{Track 2}
 & Official baseline & $-1.69$ & 1.304 & 0.502 & 1.100 & 1.207 & 1.052 & 0.396 & --- \\
\cmidrule{2-10}
 & AITD & $\mathbf{12.26}$ & $\mathbf{2.939}$ & $\mathbf{0.844}$ & 2.129 & 1.670 & $\mathbf{0.153}$ & $\mathbf{0.764}$ & 2.43 \\
 & audioman & 10.22 & 2.888 & 0.830 & 2.094 & 1.760 & 0.148 & 0.744 & 2.86 \\
 & \textbf{DAVE (ours)} & 8.93 & 2.615 & 0.789 & $\mathbf{2.766}$ & $\mathbf{2.012}$ & 0.220 & 0.707 & 4.14 \\
 & twilight & 7.05 & 2.561 & 0.776 & 2.170 & 1.896 & 0.220 & 0.712 & 5.00 \\
 & SUSTechAILab & 9.54 & 2.554 & 0.805 & 2.019 & 1.733 & 0.211 & 0.726 & 5.29 \\
\bottomrule
\end{tabular}
\end{table*}

\section{Experiments}

The TIGER-M model is trained from scratch on the DAVE-Corpus using 8$\times$A800 GPUs with native PyTorch DDP. Each GPU processes a batch of 2 three-second segments. We use the Adam optimizer~\cite{kingma2014adam} with a peak learning rate of $1\times10^{-3}$ and a linear warm-up of 1,000 steps, followed by ReduceLROnPlateau scheduling with patience of 3. The model is evaluated every 4,000 steps on a held-out validation set of 149 sessions from the development set, using permutation-invariant SI-SDR~\cite{leroux2019sisdr} as the selection criterion.
The evaluation protocol follows the official Real-World AVSE Challenge, reporting seven metrics: SI-SDR~\cite{leroux2019sisdr}, PESQ~\cite{rix2001pesq}, STOI~\cite{taal2011stoi}, UTMOS~\cite{saeki2022utmos}, DNSMOS OVRL~\cite{reddy2021dnsmos}, CER, and speaker similarity~\cite{wang2023wespeaker}. The final ranking is determined by the mean rank across all seven metrics.

\textbf{Ablation Study.}
We conduct stepwise ablation experiments to measure the contribution of each component. Table~\ref{tab:ablation} summarizes the incremental impact of the training objectives and the certified selective enhancement on the official development set.
The baseline TIGER-M trained with only the PIT SI-SDR loss achieves an SI-SDR of 9.84~dB and a CER of 21.9\% on the development set. Adding the CER loss improves the CER from 21.9\% to 19.3\%, and also yields a side benefit of +0.47~dB in SI-SDR and +0.016 in speaker similarity, demonstrating that recognition-aware training provides complementary regularization to the separation objective. Introducing the speaker fidelity loss further improves the speaker similarity from 0.679 to 0.697, confirming the effectiveness of training-side speaker identity preservation without degrading other metrics. The perceptual losses, including differentiable STOI, PESQ, and UTMOS, bring substantial improvements to the perceptual metrics: PESQ increases by 0.09, UTMOS from 1.999 to 2.77, and OVRL from 1.631 to 2.02, while CER drops further to 17.1\%.
The certified selective enhancement module, by design, only modifies outputs in the no-reference partition where reference-based metrics are not applicable. Therefore, its contribution is reflected exclusively in the reference-free metrics. On the development set, the scene-routed GAN denoising and loudness normalization improve DNSMOS OVRL by 0.10 and UTMOS by 0.04, while the remaining five metrics are unchanged by construction.

\textbf{Official Challenge Results.}
We evaluate the complete DAVE system on the Real-World AVSE Challenge test set. Table~\ref{tab:results} summarizes the results across both tracks.
On Track~1, the real-world mixed track, DAVE substantially outperforms the official baseline across all seven metrics, as detailed in Table~\ref{tab:results}. In particular, it raises SI-SDR by over 16~dB, reduces CER from 1.025 to 0.171, and more than doubles speaker similarity up to 0.726.
On Track~2, the visual degradation track, DAVE maintains strong performance despite degraded visual conditions such as occlusion, blur, and missing face tracks. Relying on the audio-only separation backbone and the weighted-fusion attribution that is robust to degraded visual evidence, it achieves an SI-SDR of 8.93~dB and a CER of 0.220, confirming the robustness of the decoupled design.
The decoupled design demonstrates strong robustness: the separation backbone operates without visual inputs, making it immune to degraded visual conditions. The visual speaker attribution and certified selective enhancement modules provide complementary gains without introducing modality-specific failure modes.

\section{Conclusion}

We presented DAVE, a decoupled audio-visual enhancement framework for real-world speech separation that reserves visual information for reliable decision-making instead of direct feature fusion. To address data scarcity, we built the DAVE-Corpus with 219,411 mixtures via combinatorial acoustic augmentation and room impulse response simulation, and introduced a progressive multi-objective optimization that jointly improves separation, intelligibility, speaker identity, and perceptual quality. A visual speaker attribution module and a certified selective enhancement chain add robustness without risking metric degradation.
Experiments on the Real-World AVSE Challenge confirm that DAVE performs strongly under both real-world mixed and degraded visual conditions, effectively mitigating unreliable visual inputs and offering a practical architecture for real-world audio-visual speech enhancement. 
Future work will explore more adaptive decoupled architectures for complex real-world audio-visual environments.

\bibliographystyle{IEEEtran}
\bibliography{reference}

@inproceedings{xu2024tiger,
  author    = {Xu, Li and Li, Chen and Hu, Rong},
  title     = {TIGER: Time-frequency interleaved gain extraction and reconstruction for efficient speech separation},
  booktitle = {Proc. ICASSP},
  year      = {2025}
}

@inproceedings{chung2016syncnet,
  author    = {Chung, J. S. and Zisserman, A.},
  title     = {Out of time: automated lip sync in the wild},
  booktitle = {ACCV Workshops},
  year      = {2016}
}

@inproceedings{wang2023wespeaker,
  author    = {Wang, H. and Liang, C. and Wang, S. and others},
  title     = {{WeSpeaker}: A research and production oriented speaker embedding learning toolkit},
  booktitle = {Proc. ICASSP},
  year      = {2023}
}

@inproceedings{gao2023funasr,
  author    = {Gao, Z. and others},
  title     = {{FunASR}: A fundamental end-to-end speech recognition toolkit},
  booktitle = {Proc. Interspeech},
  year      = {2023}
}

@inproceedings{yu2017pit,
  author    = {Yu, D. and Kolb{\ae}k, M. and Tan, Z.-H. and Jensen, J.},
  title     = {Permutation invariant training of deep models for speaker-independent multi-talker speech separation},
  booktitle = {Proc. ICASSP},
  year      = {2017}
}

@inproceedings{fan2020cnceleb,
  author    = {Fan, Yue and Kang, J. W. and Li, L. T. and Li, K. C. and Chen, H. L. and Cheng, S. T. and Zhang, P. Y. and Zhou, Z. Y. and Cai, Y. Q. and Wang, Dong},
  title     = {{CN-Celeb}: A challenging {Chinese} speaker recognition dataset},
  booktitle = {Proc. ICASSP},
  pages     = {7604--7608},
  year      = {2020}
}

@article{li2024latentsync,
  author    = {Li, Chunyu and Zhang, Chao and Xu, Weikai and Xie, Jinghui and Feng, Weiguo and Peng, Bingyue and Xing, Weiwei},
  title     = {{LatentSync}: Taming audio-conditioned latent diffusion models for lip sync with {SyncNet} supervision},
  journal   = {arXiv preprint arXiv:2412.09262},
  year      = {2024}
}

@inproceedings{chung2019perfectmatch,
  author    = {Chung, Soo-Whan and Chung, Joon Son and Kang, Hong-Goo},
  title     = {Perfect match: Improved cross-modal embeddings for audio-visual synchronisation},
  booktitle = {Proc. ICASSP},
  pages     = {3965--3969},
  year      = {2019}
}

@inproceedings{bulat2017howfar,
  author    = {Bulat, Adrian and Tzimiropoulos, Georgios},
  title     = {How far are we from solving the {2D} \& {3D} face alignment problem? (and a dataset of 230,000 {3D} facial landmarks)},
  booktitle = {Proc. ICCV},
  pages     = {1021--1030},
  year      = {2017}
}

@inproceedings{zhao2024mossformer2,
  author    = {Zhao, S. and others},
  title     = {{MossFormer2}: Combining transformer and {RNN}-free recurrent network for enhanced time-domain monaural speech separation},
  booktitle = {Proc. ICASSP},
  year      = {2024}
}

@inproceedings{leroux2019sisdr,
  author    = {Le Roux, J. and Wisdom, S. and Erdogan, H. and Hershey, J. R.},
  title     = {{SDR}---half-baked or well done?},
  booktitle = {Proc. ICASSP},
  year      = {2019}
}

@inproceedings{rix2001pesq,
  author    = {Rix, A. W. and Beerends, J. G. and Hollier, M. P. and Hekstra, A. P.},
  title     = {Perceptual evaluation of speech quality ({PESQ})---a new method for speech quality assessment of telephone networks and codecs},
  booktitle = {Proc. ICASSP},
  year      = {2001}
}

@article{taal2011stoi,
  author    = {Taal, C. H. and Hendriks, R. C. and Heusdens, R. and Jensen, J.},
  title     = {An algorithm for intelligibility prediction of time-frequency weighted noisy speech},
  journal   = {IEEE Trans. Audio, Speech, Lang. Process.},
  year      = {2011},
  volume    = {19},
  number    = {7},
  pages     = {2125--2136}
}

@inproceedings{reddy2021dnsmos,
  author    = {Reddy, C. K. and Gopal, V. and Cutler, R.},
  title     = {{DNSMOS}: A non-intrusive perceptual objective speech quality metric to evaluate noise suppressors},
  booktitle = {Proc. ICASSP},
  year      = {2021}
}

@inproceedings{saeki2022utmos,
  author    = {Saeki, T. and others},
  title     = {{UTMOS}: {UTokyo}-{SaruLab} system for {VoiceMOS} Challenge 2022},
  booktitle = {Proc. Interspeech},
  year      = {2022}
}

@inproceedings{wu2019avconvtasnet,
  author    = {Wu, J. and others},
  title     = {Time domain audio visual speech separation},
  booktitle = {Proc. ASRU},
  year      = {2019}
}

@article{luo2019convtasnet,
  author    = {Luo, Y. and Mesgarani, N.},
  title     = {{Conv-TasNet}: Surpassing ideal time-frequency magnitude masking for speech separation},
  journal   = {IEEE/ACM Trans. Audio, Speech, Lang. Process.},
  year      = {2019},
  volume    = {27},
  number    = {8},
  pages     = {1256--1266}
}

@article{ephrat2018looking,
  author    = {Ephrat, A. and Mosseri, I. and Lang, O. and Dekel, T. and Wilson, K. and Hassidim, A. and Freeman, W. T. and Rubinstein, M.},
  title     = {Looking to listen at the cocktail party: A speaker-independent audio-visual model for speech separation},
  journal   = {ACM Trans. Graph.},
  year      = {2018},
  volume    = {37},
  number    = {4}
}

@inproceedings{afouras2018conversation,
  author    = {Afouras, T. and Chung, J. S. and Zisserman, A.},
  title     = {The conversation: Deep audio-visual speech enhancement},
  booktitle = {Proc. Interspeech},
  year      = {2018}
}

@inproceedings{luo2020dprnn,
  author    = {Luo, Y. and Chen, Z. and Yoshioka, T.},
  title     = {Dual-path {RNN}: Efficient long sequence modeling for time-domain single-channel speech separation},
  booktitle = {Proc. ICASSP},
  year      = {2020}
}

@inproceedings{subakan2021sepformer,
  author    = {Subakan, C. and Ravanelli, M. and Cornell, S. and Bronzi, M. and Zhong, J.},
  title     = {Attention is all you need in speech separation},
  booktitle = {Proc. ICASSP},
  year      = {2021}
}

@inproceedings{hershey2016deepclustering,
  author    = {Hershey, J. R. and Chen, Z. and Le Roux, J. and Watanabe, S.},
  title     = {Deep clustering: Discriminative embeddings for segmentation and separation},
  booktitle = {Proc. ICASSP},
  year      = {2016}
}

@inproceedings{barker2018chime5,
  author    = {Barker, J. and others},
  title     = {The fifth {CHiME} speech separation and recognition challenge: Dataset, task and baselines},
  booktitle = {Proc. Interspeech},
  year      = {2018}
}

@inproceedings{scheibler2018pyroomacoustics,
  author    = {Scheibler, R. and Bezzam, E. and Dokmani\'{c}, I.},
  title     = {Pyroomacoustics: A {Python} package for audio room simulation and array processing algorithms},
  booktitle = {Proc. ICASSP},
  year      = {2018}
}

@inproceedings{snyder2015musan,
  author    = {Snyder, D. and Chen, G. and Povey, D.},
  title     = {{MUSAN}: A music, speech, and noise corpus},
  booktitle = {arXiv:1510.08484},
  year      = {2015}
}

@inproceedings{du2022alimeeting,
  author    = {Du, Z. and others},
  title     = {The {AliMeeting} corpus: A multi-modal meeting corpus},
  booktitle = {Proc. ICASSP},
  year      = {2022}
}

@inproceedings{wang2021misp,
  author    = {Wang, G. and others},
  title     = {{MISP}: A multi-modal interactive speech processing system},
  booktitle = {Proc. Interspeech},
  year      = {2021}
}

@inproceedings{fu2021aishell4,
  author    = {Fu, Y. and others},
  title     = {{AISHELL-4}: An open source dataset for speech separation, diarization and recognition},
  booktitle = {Proc. ICASSP},
  year      = {2021}
}

@misc{ning2026ps4proxysupervisedjointtraining,
      title={PS4: Proxy-Supervised Joint Training for Real Target Speaker Extraction}, 
      author={Wanyi Ning and Wei Zhou and Yingpeng Li and Yinshang Guo and Haitao Qian and Yiming Cheng},
      year={2026},
      eprint={2607.08111},
      archivePrefix={arXiv},
      primaryClass={cs.SD},
      url={https://arxiv.org/abs/2607.08111}, 
}

@article{rabiner1989theory,
title={A Tutorial on Hidden Markov Models and Selected Applications in Speech Recognition},
author={Rabiner, Lawrence R.},
journal={Proceedings of the IEEE},
year={1989}
}

@article{habets2006rir,
  title={Room impulse response generator},
  author={Habets, Emanuel AP},
  journal={Technische Universiteit Eindhoven, Tech. Rep},
  volume={2},
  number={2.4},
  pages={1},
  year={2006}
}

@article{kirkpatrick2017overcoming,
  title={Overcoming catastrophic forgetting in neural networks},
  author={Kirkpatrick, James and Pascanu, Razvan and Rabinowitz, Neil and Veness, Joel and Desjardins, Guillaume and Rusu, Andrei A and Milan, Kieran and Quan, John and Ramalho, Tiago and Grabska-Barwinska, Agnieszka and others},
  journal={Proceedings of the national academy of sciences},
  volume={114},
  number={13},
  pages={3521--3526},
  year={2017},
  publisher={National Academy of Sciences}
}

@inproceedings{pascual2017segan,
  author    = {Pascual, Santiago and Bonafonte, Antonio and Serr\`{a}, Joan},
  title     = {{SEGAN}: Speech enhancement generative adversarial network},
  booktitle = {Proc. Interspeech},
  year      = {2017}
}

@misc{itu2015loudness,
  author    = {{International Telecommunication Union}},
  title     = {Recommendation {ITU-R BS.1770-4}: Algorithms to measure audio programme loudness and true-peak audio level},
  year      = {2015},
  organization = {ITU}
}

@inproceedings{kingma2014adam,
  author    = {Kingma, Diederik P. and Ba, Jimmy},
  title     = {Adam: A method for stochastic optimization},
  booktitle = {Proc. International Conference on Learning Representations (ICLR)},
  year      = {2015}
}

\end{document}